\documentclass[reprint,aps,prl,twocolumn,amsmath,amssymb,footinbib,superscriptaddress]{revtex4-2}
\usepackage{graphicx,xcolor}
\usepackage{epstopdf}
\usepackage{xr-hyper}
\usepackage{hyperref}
\usepackage{cleveref}

\newcommand{\gwg}{\textsf{z}}
\newcommand{\loss}{\Gamma}

\newcommand{\dn}{\mathrm{dn}}
\newcommand{\vS}{c_S}

\begin{document}

\title{
Microwave Circulation in an Extended Josephson Junction Ring }

\author{Dat Thanh Le} \email[]{Contact author: dat.le@aqcircuits.com}
\affiliation{Analog Quantum Circuits Pty. Ltd., Ipswich, Queensland 4300, Australia}

\author{Arkady Fedorov}
\affiliation{Analog Quantum Circuits Pty. Ltd., Ipswich, Queensland 4300, Australia}
\affiliation{School of Mathematics and Physics,
University of Queensland, Brisbane, Queensland 4072, Australia}

\author{T. M. Stace} \email[]{Contact author: tom@aqcircuits.com}
\affiliation{Analog Quantum Circuits Pty. Ltd., Ipswich, Queensland 4300, Australia}

\begin{abstract}
Circulators are nonreciprocal devices that enable directional signal routing.  Nonreciprocity, which requires time-reversal symmetry breaking, can be produced in waveguides in which the propagation medium moves relative to the waveguide at a moderate fraction of the wave speed.  Motivated by this effect, here we propose a design for nonreciprocal microwave transmission based on an extended, annular Josephson junction, in which the propagation medium consists of a train of moving fluxons. 
We show how to harness this  to build a high-quality resonant microwave circulator, and we  theoretically evaluate the anticipated performance of such a device. 

\end{abstract}

\maketitle
 
Electromagnetic wave propagation is typically reciprocal, in which scattering is invariant under the exchange of the source and the receiver \cite{PozarBook11}.  However, breaking reciprocity is essential for realising nonreciprocal devices such as isolators and circulators that find  applications in numerous fields \cite{Caloz18}, including quantum technologies based on superconducting circuits. 

Conventional microwave circulators are bulky, ferrite-based, off-devices which complicates their use with sensitive superconducting microelectronics. Designs for compact, non-ferritic, on-chip circulators are of great interest for superconducting quantum technologies, and include active AC-driven (or cyclically-switched) interferometers \cite{Kamal11,Sliwa15,Lecocq17,PhysRevApplied.22.064020}, and passive DC-controlled, flux-biased, SQUID-like loops of  Josephson junctions \cite{Koch10,Muller18,Le21,Navarathna23,Fedorov24,Kumar25}.  However AC-control introduces design challenges, while flux-biased junction loops are limited by charge noise and  low saturation power.   

There has been a recent flourishing in alternative approaches to induce nonreciprocal  propagation in dynamical systems \cite{ShenPRL23,ZhangPRL24,veenstra2024non,WangPRL24}, and formal approaches to quantising non-reciprocal circuits \cite{RodriguezQuantum24,LabarcaPRAp24,RodriguezPRX25,RymarzPRX23,rymarz2018quantum,ParraRodriguezPRB19,EgusquizaPRB22,RymarzPRX21,
ParraRodriguez2022canonical,rymarz2023nonreciprocal}. 
 One unifying mechanism is to set a waveguide medium in motion relative to the transmitter and receiver \cite{PhysRevLett.93.126804,Fleury14,PhysRevX.4.021019,Mahoney17,HuangPRL18}.  
For example, \citet{Fleury14} fan-forced air around a ring-resonator at a few percent of the speed-of-sound  to create a 3-port acoustic circulator.  

Here we propose a method to use the effective motion of a synthetic electronic  medium to realise an on-chip, DC-controlled, superconducting microwave circulator. Our proposal is based on an annular long Josephson junction (LJJ), an extended nonlinear system which supports microwave transmission in its soliton excitations, known as \emph{fluxons} \cite{Kulik67}.  For nominally stationary fluxons, transmission occurs via pairwise-degenerate counter-propagating  modes. This degeneracy is lifted by forcing fluxons into a bulk motional state using an applied DC bias current,  thereby breaking time-reversal symmetry. This yields
a strongly nonreciprocal microwave response, forming the basis of our circulator proposal.  

\begin{figure}[t!]
    \raggedright
    \includegraphics[width=0.95\columnwidth]{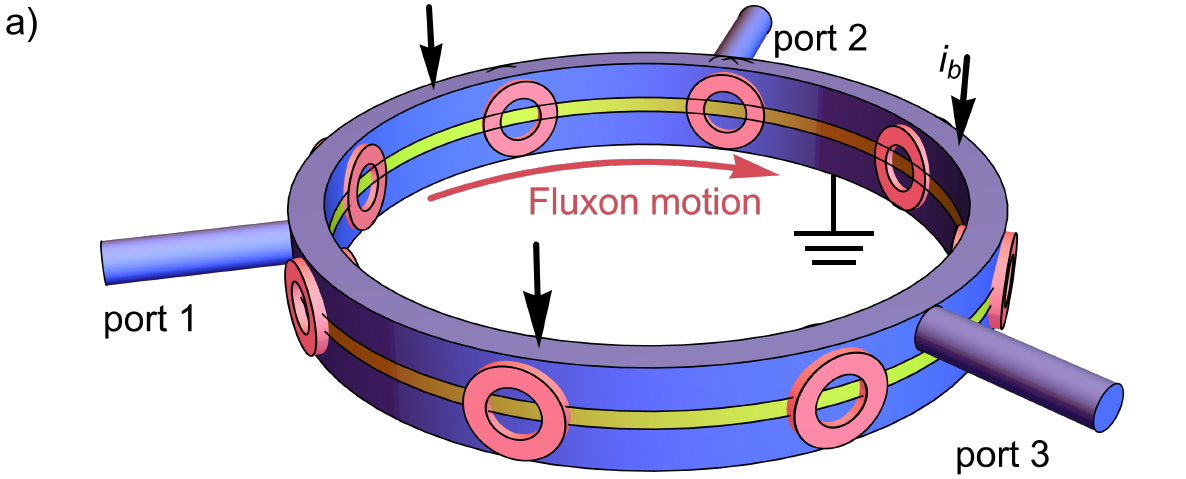}\hfill
    \vspace{2mm}
    \includegraphics{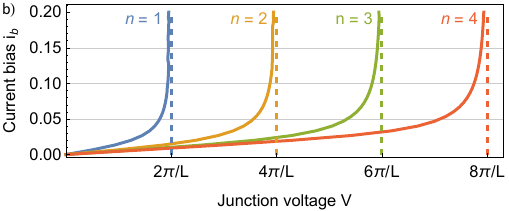}
    \caption{(a) Cartoon of an annular LJJ (not to scale), consisting of two  superconducting rings (blue) formed from evaporated aluminium films, and separated by a  tunnel-barrier junction (yellow), and coupled  to three external ports, depicting the central conductors of the port waveguides as cylinders. The $n=8$  `wagon wheels'  represent the current density loops associated to localised fluxons  (pink).  Vertical arrows indicate a current bias $i_b$ through the junction to induce fluxon motion. (b) Dimensionless DC I-V response of an annular LJJ for different numbers of fluxons $n$, with  system parameters $\textsf{L}=15$ and $\textsf{g}=0.02$. Dashed lines represent the asymptotic voltage $\textsf{V}=2\pi n/\textsf L$, determined by the Swihart velocity.}
    \label{fig:LongJunction}
\end{figure}

We briefly revisit the physics of LJJs, and then establish their utility for resonant microwave circulation. 
An annular  LJJ depicted in \cref{fig:LongJunction}a is a spatially extended Josephson junction, whose length $L$, width $w$, and  Josephson penetration depth $\lambda_J$ satisfy $w\ll\lambda_J\ll L$ \cite{Mazo14}.  The superconducting junction phase $\phi(x,t)$ varies along the long axis, $0<x<L$. The dynamics of $\phi$ is governed by the sine-Gordon (SG) equation \cite{McLaughlin78}, 
whose  soliton solutions in a LJJ correspond to localised current loops (fluxons)  circulating through the junction, each with quantised flux  $\Phi_0=h/(2e)$. 


An applied current bias $i_b$ through the junction induces fluxons to move along the LJJ with velocity $v$, which is tuneable from $v=0$ up to a terminal velocity given by the Swihart velocity,  \mbox{$\vS$}. Typically, \mbox{$\vS/c\sim10^{-2}$}, where $c$ is the vacuum speed-of-light \cite{Swihart61}. Dissipation of fluxon motion  in a LJJ is low
 \cite{Wallraff03}, unlike Abrikosov vortices in a type-II superconductor \cite{TinkhamBook2004} and fluxons in a discrete junction array \cite{Fazio01}.  

Fluxon trains in a LJJ support propagating microwaves with group-velocity $\sim\vS$ \cite{Kulik67}, making it the effective speed-of-light in a LJJ transmission line. As a result, the fluxon speed can be tuned from zero to extremely relativistic.  We use this effect to produce  nonreciprocal microwave scattering. In particular, moving fluxons  may support or attenuate microwave propagation, depending on the propagation direction  relative to the fluxon flow  \cite{Fedorov12, Pankratov15}, analogous to travelling-wave optical isolators \cite{Maayani18}. 

In what follows, we consider a moving fluxon train in an annular LJJ  \cite{Davidson85,Ustinov92Third, Vernik92, Ustinov02}, coupled to external microwave ports, illustrated in \cref{fig:LongJunction}a.  We compute the microwave response of this system, and show that it  exhibits  circulation near the resonance frequency \mbox{$f_{\rm res}\approx\vS/L$}. Since $\vS\ll c$, this  allows for a compact on-chip microwave circulator.

The non-dimensionalised SG equation governing the LJJ dynamics is \cite{McLaughlin78}
\begin{equation}
    \phi_{\textsf{tt}} - \phi_{\textsf{xx}} + \sin (\phi) = \textsf{i}_b - \textsf{g} \phi_{\textsf{t}} + \textsf p \phi_{\textsf{xxt}}+\mathcal{C}_{\rm x}[\phi], \label{eq:SGEquation}
\end{equation}
with non-dimensionalised space, $\textsf{x}\equiv x/\lambda_J $, and time, \mbox{$\textsf{t} \equiv t \omega_p$},  and plasma frequency \mbox{$\omega_p = \vS/\lambda_J=2\pi f_p$}.   Subscripted $\textsf{x}$ or $\textsf{t}$ coordinates denote multi-order partial derivatives, $\mathcal{C}_{\rm x}$ represents  external waveguide coupling terms (see supplementary information \cite{suppinfo} and Refs.\ \cite{fedorov2013fluxon,MeckbachPhdThesis2013}), $\textsf{i}_b$ is a uniform current bias, and $\textsf{g}$ and $\textsf{p}$ are dimensionless temperature-dependent parameters quantifying quasiparticle dissipation and surface loss. Typically, $\textsf p \ll \textsf g$, so in the following analysis we set $\textsf p=0$. In an annular geometry with perimeter $L$, $\phi$ satisfies the periodicity condition $ \phi(\textsf{x}+ \textsf{L}) = \phi(\textsf{x}) + 2\pi n$, where $\textsf{L}\equiv L/\lambda_J$ and $n\in \mathbb{Z}$ is the number of fluxons.

For an uncoupled LJJ  (i.e.\ $\mathcal{C}_{\rm x}=0$) the Lorentz-invariant  solution to the SG \cref{eq:SGEquation} is 
\begin{equation}
     \phi_0 (\textsf{x},\textsf{t}) = \pi + 2\, \mathrm{am} \left( {\gamma_\textsf{v} (\textsf{x} - \textsf{vt})}/{k}, k \right),  \label{eq:SteadyFluxonSolution}
\end{equation}
where $\mathrm{am}(u,k)$ is the Jacobi amplitude function,  $\textsf{v}=v/\vS$ is the dimensionless fluxon velocity, and \mbox{$\gamma_\textsf{v} = (1-\textsf{v}^2)^{-1/2}$} is the Lorentz factor \cite{Malomed90}. The  Jacobi modulus $k$ and $\textsf{v}$ depend implicitly on the fluxon spacing $\textsf{L}/n$ and $\textsf{i}_b$  through   \cite{Shnirman94}
\begin{align}
	\textsf{L}/n&= 2 k K(k)/\gamma_\textsf{v}, \label{eq:ModulusConditionForMovingFluxons}\\
	\textsf{i}_b &= - {4 \gamma_\textsf{v}  \textsf{v}\,\textsf{g} E(k)}/{(\pi k)}  , \label{eq:CurrentDissipationRelation}
\end{align}
where $K(k)$ and  $E(k)$ are the complete elliptic integrals of the first- and  second-kind. 

A moving fluxon train generates a local voltage, \mbox{$V\equiv\Phi_t=\Phi_0 f_p \textsf{V}$} where \mbox{$\textsf{V}=\phi_\textsf{t}$}.  The time-averaged DC contribution to the voltage  is $\textsf{V}_{\rm DC} = {2\pi n}\textsf{v}/\textsf{L}=- {\pi^2 \textsf{i}_b}/({4 \textsf{g} E(k) K(k)})$.  If $\textsf{L}/n$ and $\textsf{v}$ are both small, then  $\textsf{V}_{\rm DC} \approx - \textsf{i}_b/\textsf{g}$. 
\Cref{fig:LongJunction}b shows characteristic \mbox{I-V} curves for different fluxon numbers, with $\textsf{L}=10$ and $\textsf{g}=0.02$, showing the linear response for low voltage, and the asymptote $\textsf{V}_{\rm DC}=2\pi n/\textsf L$ at large $\textsf{i}_b$ where ${\textsf{v}\rightarrow1}$ \cite{Ustinov02}.

We analyse microwave propagation perturbatively around $\phi_0$, 
setting  \mbox{$ \phi(\textsf{x},\textsf{t}) = \phi_0 (\textsf{x},\textsf{t}) + \psi(\textsf{x},\textsf{t})$}. 
For stationary fluxons ($\textsf{v}=0$), and linearising \cref{eq:SGEquation} in $\psi$ yields Lame's equation  \cite{WhittakerAndWatson89} 
\begin{equation}
    \psi_{\textsf{tt}} - \psi_{\textsf{xx}} + \cos (\phi_0(\textsf{x})) \psi = 0. \label{eq:OscillationEquation}
\end{equation}
The periodicity \mbox{$\cos(\phi_0(\textsf{x}+\textsf{L}/n)) = \cos (\phi_0 (\textsf{x}))$} yields 
Bloch solutions   \mbox{$ \psi(\textsf{x},\textsf{t}) =  e^{i q \textsf{x}}  u(\textsf{x}) e^{\pm i \omega \textsf{t}}$}, where $q$ is the wavenumber, \mbox{$u(\textsf{x})=u(\textsf{x}+\textsf{L}/n)$}, 
and $\omega=2\pi f/\omega_p$ is the dimensionless eigenfrequency. 
Explicitly, $u(\textsf{x})$, $q$, and $\omega$ are given by \cite{Shnirman94}
\begin{align}
 u(\textsf{x}) &=e^{ \pm i {\pi \textsf{x}}/{(2 k K(k)) }} {H(\textsf{x}/k \pm \beta )}/{\Theta (\textsf{x}/k)}, \label{eq:LameSolution} \\
 q &= \mp \big(  {Z(\beta)}/{(ik)}+{\pi}/{(2 k K(k))} \big), \label{eq:LameWavenumber} \\
 \omega &= {\dn (\beta)}/{k}, \label{eq:LameFrequency}
\end{align}
where $H$, $\Theta$, and $Z$ are  Jacobi's eta, theta, and zeta functions, respectively, $\dn$ is a Jacobi elliptic function, and $\beta$ is a complex parameter. The $\pm q$  solutions of  \cref{eq:LameWavenumber} represent  two-fold degenerate counter-propagating modes.

\begin{figure}[t!]
    \centering
    \includegraphics{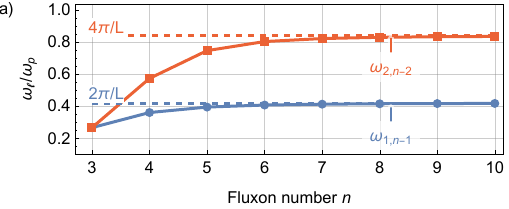}
    \includegraphics{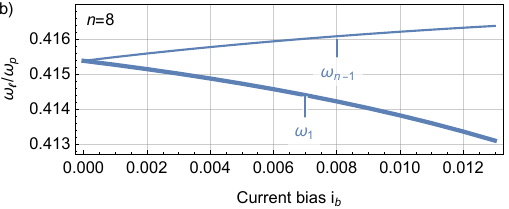}
    \caption{(a) 
    Pairwise-degenerate fluxon-mode frequencies $\omega_1,\omega_{n-1}$ and $\omega_2,\omega_{n-2}$
    (markers) versus the fluxon number $n$, with the asymptote (dashed), shown for \mbox{$\textsf{i}_b=0$}, \mbox{$\textsf{L}=15$}.  (b)  The degeneracy of the counter-rotating modes $\omega_1$ and $\omega_{n-1}$ is lifted for non-zero current bias $\textsf{i}_b$, shown for  $n=8$,  
     $\textsf{g}=0.02$.  }
    \label{fig:FluxonOscillationFrequencies}
\end{figure}

Roots of \cref{eq:LameFrequency} are characterised by two frequency bands separated by a spectral gap \cite{Lebwohl67,Fetter68}: (I) the fluxon band  \mbox{$ 0 \leq \omega^2 \leq 1/k^2-1$} with ${\rm re}(\beta)=K(k)$, and  (II) the plasma band \mbox{$\omega^2\geq {1}/{k^2} $}  with ${\rm re}(\beta)=0$. In both bands, ${\rm im}(\beta)\in [0,K(\sqrt{1-k^2})]$.
For the ring geometry, \mbox{$\psi(\textsf{x}+\textsf{L},\textsf{t}) = \psi(\textsf{x},\textsf{t})$} requires that $q \textsf{L} /( 2\pi)\equiv \ell \in \mathbb{Z}$, and from \cref{eq:LameWavenumber} we find that $\beta$ is quantised, and is a root of 
\begin{equation}
    Z(\beta) = {i  (2\ell - n)\pi}/({2n K(k)}), %
    \label{eq:betaCondition}
\end{equation}
where $Z(\beta)$ is purely imaginary.
For $\beta$ in the fluxon band (I), we find that \mbox{${|Z(\beta)|}\leq  \pi / (2K(k))$}, so that \mbox{$\ell\in\{0,...,n\}$} %
\cite{Shnirman94}. These $n+1$ discrete modes are two-fold degenerate, with $\omega_{\ell} = \omega_{n-\ell}$, corresponding to the counter-propagating degenerate modes.  In addition, $\omega_0 = \omega_n = 0$.  %
We show $\omega_{\ell}$ as functions of the fluxon number $n$ in \cref{fig:FluxonOscillationFrequencies}a, illustrating that for large $n$, $\omega_\ell\lesssim 2\pi \ell /\textsf{L}$. %
This gives  characteristic operating frequencies in the microwave band.

\begin{figure}[t!]
    \includegraphics{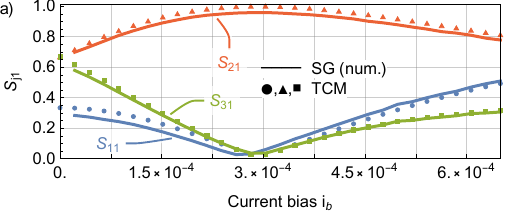}
     \includegraphics{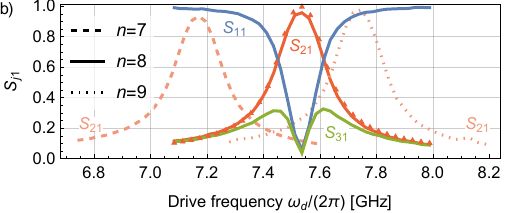}
    \caption{
    (a)
    $|S_{j1}|$ dependence on the dimensionless current bias $\textsf{i}_b$ at the resonant frequency $\omega_d/(2\pi) = 7.53 \, \mathrm{GHz}$; $|S_{21}|$ is maximised at $\textsf{i}_b = 3 \times 10^{-4}$. Solid lines are from numerical solutions of the input-output SG \Cref{eq:SGEquation}, and markers are from the temporal coupled-mode (TCM) model.
     (b) 
     $|S_{j1}|$ dependence  on the drive frequency $\omega_d$ at $\textsf{i}_b = 3 \times 10^{-4}$ at $n=8$ (solid lines).  Also shown is   
     $|S_{21}|$ for $n=7,9$ (dashed and dotted lines).   In both panels $\textsf{L}=26$.}
    \label{fig:SmatrixAnalysis}
\end{figure}

For moving fluxons,  the quantisation condition is   \cite{Shnirman94}
\begin{equation}
    Z(\beta) + i \omega \textsf{v} k = {i  (2\ell-n)\pi}/({2n K(k)}). \label{eq:betaConditionRotatingFrame}
\end{equation}
The velocity dependence in \cref{eq:betaConditionRotatingFrame}  lifts the degeneracy of $\omega_{\ell} $ and $ \omega_{n-\ell}$, which breaks time-reversal symmetry.  This  underpins the circulator proposed here \cite{PozarBook11,Fleury14}.
  We illustrate the frequency splitting in \cref{fig:FluxonOscillationFrequencies}b, showing $\omega_1$ and $\omega_{n-1}$ as functions of the  current bias $\textsf{i}_b$. %

We analyse microwave scattering from an annular LJJ coupled to three external ports positioned symmetrically around the junction ring, as shown in \cref{fig:LongJunction}a. Using the input-output theory for galvanic coupling \cite{PozarBook11,ClerkRMP10}, the external  waveguide coupling term in \cref{eq:SGEquation} is
\mbox{$
\mathcal{C}_{\rm x}[\phi]={\sum}_{j=1}^3 {\gwg} \delta (\textsf{x} -\textsf{x}_j) (2 \textsf{V}_{\text{in}}^{(j)}(\textsf t) - \phi_{\textsf{t}})
$}, where $\textsf{x}_j$ are the port locations, and %
$\gwg = Z_{\mathrm{LJJ}}/Z_0 $  is the  coupling strength between the LJJ and the waveguides, whose impedances are $Z_{\mathrm{LJJ}}$ and $Z_0$.  The  input-output relation at  port $j$ is
\begin{align}
 \phi_{\textsf{t}} (\textsf{x}_j,\textsf{t}) &= \textsf{V}_{\mathrm{in}}^{(j)}(\textsf{t}) + \textsf{V}_{\mathrm{out}}^{(j)}(\textsf{t}), \label{eq:InputOutputRelation}
\end{align}
where $\textsf{V}_{\mathrm{in/out}}^{(j)}$ are the input/output voltages.
To compute the scattering matrix  $S$ at a drive frequency $\omega_d$, we solve equations (\ref{eq:SGEquation}) and (\ref{eq:InputOutputRelation}) numerically with \mbox{$\textsf{V}_{\mathrm{in}}^{(k)}(\textsf t)={\tilde{\textsf{V}}}_{\mathrm{in}}^{(k)} \cos (\omega_d \textsf t)$}, and then compute the  output Fourier amplitude ${{\tilde{\textsf{V}}}_{\mathrm{out}}^{(j)}}$ at  $\omega_d$, from which we derive the scattering matrix element \mbox{$S_{jk} ={{\tilde{\textsf{V}}}_{\mathrm{out}}^{(j)}}/{\tilde{\textsf{V}}}_{\mathrm{in}}^{(k)}  $ }\cite{PozarBook11,Fedorov24}. %

\begin{figure}[t]
    \centering
    \includegraphics{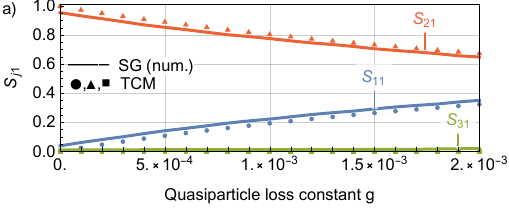} 
    \includegraphics{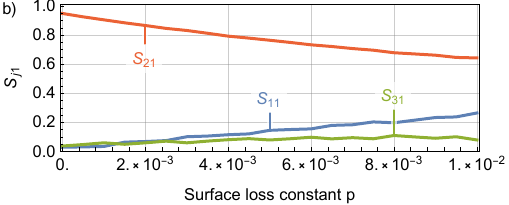}
    \caption{ (a) Dependence of $S$-matrix elements  on the dimensionless quasiparticle loss, $\textsf g$, computed  numerically  from  the  SG \Cref{eq:SGEquation}  (solid lines), and  TCM (markers). (b) Dependence of $S_{j1}$  on the dimensionless surface loss,  $\textsf p$.  In both panels, the external coupling is $\loss_{\rm x}=0.002$.
    }
    \label{fig:VaryingSystemDissipation}
\end{figure}

In what follows, for numerical simulations we assume  parameter values $\lambda_J  = 37.9\, \mu\mathrm{m}$,  \mbox{$\omega_p/(2\pi)= 33.5\, \mathrm{GHz}  $}, \mbox{$\vS = 7.98\times 10^{6}\, \mathrm{m/s}= 0.027 c$}, $Z_{\mathrm{LJJ}}= 1.82\, \Omega$, and \mbox{$Z_0= 50\, \Omega$}  \cite{Abdumalikov05}. Based on an analysis similar to that in \cref{fig:FluxonOscillationFrequencies}a, we choose $\textsf{L}=26$ and $n=8$ so that the resonant frequency is $f_{\rm res}=\omega_1/(2\pi)=7.53 \textrm{ GHz}$.

To illustrate the proposed circulator performance, in \cref{fig:SmatrixAnalysis}a, we fix the drive frequency at the fluxon mode resonance $\omega_d=\omega_1$, %
 and show $|S_{j1}|$ as functions of the current bias $\textsf{i}_b$. We see that $S_{21}$ peaks near unity at $\textsf{i}_b\approx3\times 10^{-4}$, where  $S_{11}$ and $S_{31}$ are correspondingly small. At the optimal current bias, the dimensionless fluxon velocity is $\textsf{v}\approx 0.036$. %

\Cref{fig:SmatrixAnalysis}b shows  $|S_{j1}|$ as functions of $\omega_d$ near resonance, for $n=8$ (solid lines), using the optimal $\textsf{i}_b$ from \cref{fig:SmatrixAnalysis}a.  %
For the parameter choice used here, the bandwidth at full-width--half-maximum is  $B\approx 210\, \mathrm{MHz}$. %
From the numerical SG simulations, we also compute the time-averaged voltage $V_{\rm DC}\approx 4.8 \, \mu\mathrm{V}$, generated by fluxons moving around the  ring. The three waveguides provide an external load of $Z_0/3$, which implies power dissipation of  $P_{\rm diss}\approx 1.5\textrm{ pW}=-88\textrm{ dBm}$. %

In addition, since the resonant frequency of the  LJJ ring depends on the number of fluxons, it is possible to  tune the resonant circulation frequency \emph{in-situ}. This is also illustrated in \cref{fig:SmatrixAnalysis}b,  which shows the forward circulation amplitude, $S_{21}$, for $n=7$ and 9 fluxons. additional analysis of the  dependence of $S_{21}$ on $n$ is provided in the supplementary material \cite{suppinfo}. %

These numerical predictions are replicated with a simple temporally coupled-mode (TCM) scattering model \cite{Haus1984Book,Fan03,Zhao19}.  To compute the $S$-matrix, TCM assumes  two counter-propagating modes at frequencies $\omega_\pm$ with detuning $\Delta\omega= \omega_+ - \omega_- $,  coupling equally to the  external waveguides with coupling strength $\loss_{\rm x}$ \cite{Wang05}.  Assuming no internal losses, the forward circulation amplitude is
\begin{equation}
 S_{21}  =   -\tfrac{2}{3} {\sum}_{\sigma=\pm} {\loss_{\rm x} e^{\sigma i \pi/3}}/\big({\loss_{\rm x} + i (\omega_d - \omega_\sigma)}\big). \label{eq:S21_TCM}
\end{equation}
TCM predicts  optimal circulation  when $\Delta\omega = 2 \loss_{\rm x}/\sqrt{3}$, and gives the (dimensionless) scattering bandwidth \mbox{$ \textsf{B}=3 \loss_{\rm x} $} \cite{Fleury14}.

For the  LJJ ring, we take \mbox{$\omega_+=\omega_{n-1}$}, \mbox{$\omega_-=\omega_{1}$} and  \mbox{$ \loss_{\rm x} = 3\gwg/(2\textsf{L})$}. %
\Cref{fig:SmatrixAnalysis} includes  the results of the  TCM model, %
showing reasonable agreement with  the full  numerical solution of the  SG  \Cref{eq:SGEquation}. 
Optimal circulation in the LJJ ring requires \mbox{$\Delta\omega= \sqrt{3} \gwg/ \textsf L$}.  This implicitly determines the optimal bias current, which controls $\Delta\omega$, as seen in \cref{fig:FluxonOscillationFrequencies}b. 
For the  parameters used here, TCM yields a  bandwidth of \mbox{$ B=\textsf{B} f_p\approx 210\, \mathrm{MHz} $},   consistent with  simulation results in \cref{fig:SmatrixAnalysis}b.

 \begin{figure}[t]
    \centering

    \includegraphics{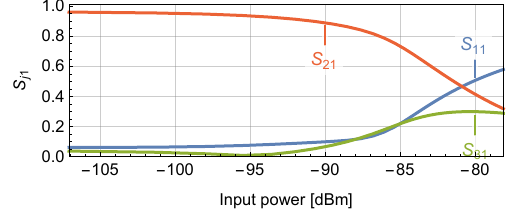} 
    \caption{%
    Input power dependence of the $S$-matrix elements $|S_{j1}|$, showing the reduction in circulation performance with increasing power.  The 1-dB power compression point occurs at $P_{\mathrm{{1\,dB}}} = -90 \, \mathrm{dBm}$.  Parameters are as in  \cref{fig:SmatrixAnalysis}. %
    }
    \label{fig:PowerAnalysis}
\end{figure}
 
 Including internal losses in the system, the total loss becomes \mbox{$ \loss_{\rm \Sigma} = \loss_{\rm x} + \loss_{\rm i}$}.  For example, for quasiparticle dissipation,  \mbox{$\loss_{\rm i}=\textsf{g}/2$}. The TCM model predicts the maximum forward scattering amplitude (at  resonance  \mbox{$\Delta\omega = 2 \loss_{\rm \Sigma} /\sqrt{3}$}), is \mbox{$|S_{21}^{\rm res}|=  {\loss_{\rm x}}/{\loss_{\rm \Sigma} }<1$}, and the reflection minimum  \mbox{$|S_{11}^{\rm res}| = 1 - {\loss_{\rm x}}/{\loss_{\rm \Sigma} }>0$},  so that circulation degrades with internal loss.

 \Cref{fig:VaryingSystemDissipation}a  illustrates the effect of quasiparticle loss,  %
showing that $S_{21}$ degrades significantly when $\loss_{\rm i} \gtrsim  \loss_{\rm x}$, as expected. {Empirically, for Nb-$\rm{AlO}_x$-Nb junctions at 4.2 K, ${\textsf g}_\textrm{4.2K}\approx0.02$ \cite{Fedorov13}. Below the critical temperature, $T_c$, thermal quasiparticles will be suppressed by a factor \mbox{$\sim \exp(-T_c/T)$}, so that  %
   ${\textsf g}_\textrm{20mK}$ is expected to be exponentially small when $T\ll T_c$. } In any specific implementation, internal and external-coupler losses could be characterised through the depinning and retrapping currents junction's IV response \cite{JohnsonPRL90}.
 
\Cref{fig:VaryingSystemDissipation}b illustrates the effect of surface loss.  {Empirically, for  Nb-$\rm{AlO}_x$-Nb junctions at 4.2 K,  ${\textsf p}_\textrm{4.2K}\approx0.03$ \cite{Fedorov12}, and since ${\textsf p}\propto (T/T_c)^{8/3}$ for  $T\ll T_c$ %
   \cite{scott1964distributed}; we estimate \mbox{$\textsf{p}_{\rm{20mK}} \sim 10^{-8}$}, which is negligibly small.}

Power saturation of the proposed LJJ circulator will occur when, for sufficiently large input signal amplitude, the moving fluxon train is strongly perturbed.  We  compute the $S$-matrix numerically at the optimal circulation point for increasing input power \mbox{{$P_{\mathrm{in}}=|{\tilde{\textsf{V}}}_{\mathrm{in}} \Phi_0 f_p|^2/Z_0 $}}, shown in \cref{fig:PowerAnalysis} (assuming the same parameters listed earlier), and find the 1-dB compression point (where $S_{21}=0.89$) is \mbox{$P_{\textrm{{1dB}}} = -90 \, \mathrm{dBm}$}.  This is much larger than the `single-photon flux' power $P_{\mathrm{photon}} = \hbar \omega_d B/ (2\pi) \approx -127\, \text{dBm} $, which sets the saturation power in other DC-controlled circulator proposals \cite{Le21,Fedorov24}.

We have proposed a microwave circulator based on synthetic-motion-induced nonreciprocity in a nonlinear  waveguide. The relatively slow microwave propagation speed in the superconducting system allows for integrated, compact, on-chip circulators, which will become increasingly important in addressing the hardware scale-up challenge in superconducting quantum technologies \cite{VirtanenPRL24,upadhyay2024microwave}.  
 We conclude by noting that annular LJJs are fabricated using standard materials and methods, allowing for  practical implementation of the proposed circulator. 
In addition, techniques for controllable insertion of fluxons into an annular LJJ \cite{Ustinov02} have been realised experimentally using DC current injection \cite{Malomed04,GaberPRB05}. Impedance engineering of the LJJ (e.g.\ to better match to external waveguides) can be achieved using novel fabrication techniques \cite{WildermuthAPL22}.

\begin{acknowledgments}

\textit{Acknowledgements}—The authors each declare a  financial interest in Analog Quantum Circuits Pty.\ Ltd.

\textit{Data availability}—The data that support the findings of this article are not publicly available because they contain commercially sensitive information. The data are available from the authors upon reasonable request.

\end{acknowledgments}

\bibliography{bib}

\clearpage

\twocolumngrid

\begin{center}
    {\Large \textbf{Supplementary Material}}
\end{center}


\setcounter{secnumdepth}{3}
\renewcommand\thesection{\Roman{section}}


\section{Derivation of the sine-Gordon equation} \label{append:SG_derivation}

In this section, we first derive the equation of motion (EOM) for a simple parallel LC circuit coupled either galvanically or capacitively to a waveguide port. This helps familiarise ourselves with calculations relevant to galvanic and capacitive couplings. We then derive the sine-Gordon (SG) equation of a long Josephson junction (LJJ) in the absence and presence of waveguide couplings.  

\subsection{Classical input-output theory: a parallel LC circuit coupled to a waveguide}

\begin{figure}[htb]
    \centering
    \includegraphics[width=0.45\textwidth]{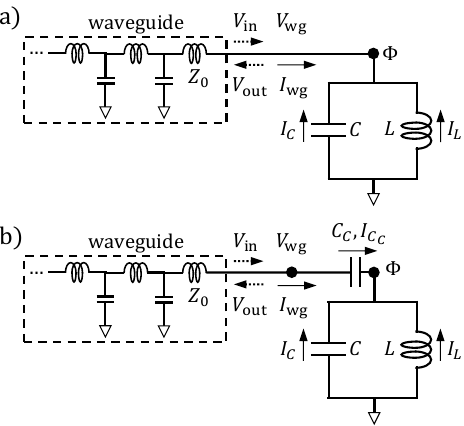}
    \caption{a) Parallel LC circuit coupled galvanically to a waveguide with an impedance $Z_0$. The waveguide is modelled in a lumped-element representation consisting of unit cells of inductors and capacitors. $V_{\rm wg}$ denotes the voltage at the waveguide boundary and is the sum of the incoming and outgoing voltages $V_{\mathrm{in}}$ and $V_{\mathrm{out}}$. 
    $I_{\rm wg}$ is the current that flows through the waveguide boundary, $I_C$ through the LC capacitor, and $I_L$ through the LC inductor. b) Same as panel a) but the parallel LC circuit is coupled capacitively to the waveguide via a coupling capacitor $C_C$.  }
    \label{fig:LCCoupledToWaveguide}
\end{figure}

We consider a parallel LC circuit coupled either galvanically or capacitively to a waveguide, as depicted in \cref{fig:LCCoupledToWaveguide}.  Given an input voltage $V_{\mathrm{in}}$ incident to the circuit from the waveguide, we compute the output voltage $V_{\mathrm{out}}$ and the reflection coefficient $\Gamma = V_{\rm out}/V_{\rm in}$.

We note here an important result regarding the voltage and current at the end (boundary) of a (lossless) waveguide \cite{PozarBook11,ClerkRMP10}
\begin{eqnarray}
V_{\rm wg} &=& V_{\mathrm{in}} + V_{\mathrm{out}} , \label{eq:waveguide_voltage} \\
I_{\rm wg} &=& \frac{V_{\rm wg} - 2 V_{\mathrm{in}}}{Z_0} . \label{eq:waveguide_current}
\end{eqnarray}
These equations will be used repeatedly when deriving circuit EOMs. 

\subsubsection{Galvanic coupling} 

We consider the circuit in \cref{fig:LCCoupledToWaveguide}a, where a parallel LC circuit is coupled galvanically to a waveguide. The reflection coefficient in this circuit can be computed via a load impedance analysis. In particular, its load impedance is
\begin{equation}
    Z_{\rm L} \equiv Z_{LC} =  \frac{i \omega L}{1- \omega^2 L C}.
\end{equation}
The reflection coefficient following the model of a waveguide of impedance $Z_0$ terminated by a load of impedance $Z_{\rm L}$ is given by \cite{PozarBook11,ClerkRMP10}
\begin{equation}
    \Gamma \equiv \frac{Z_{\rm L} - Z_0 }{ Z_{\rm L} + Z_0  } = -1 + \frac{2 i \omega L }{\omega^2 L C Z_0 - i \omega L - Z_0}. \label{eq:ReflectionCoefficientGalvanicCoupling}
\end{equation}

An alternative way to compute $\Gamma$ is using the EOM of the circuit in \cref{fig:LCCoupledToWaveguide}a. Considering the flux node $\Phi$ as denoted in the circuit, Kirchhoff's current law yields 
\begin{equation}
    I_C + I_L   + I_{\rm wg} = 0 , \label{eq:CurrentNodeEquationDirectCoupling}
\end{equation}
where $I_C = C \ddot \Phi$ and $I_L = \Phi/L$. For the circuit in \cref{fig:LCCoupledToWaveguide}a, $V_{\rm wg} \equiv \dot \Phi$, so following eqs.\ \eqref{eq:waveguide_voltage} and \eqref{eq:waveguide_current} we have
\begin{equation}
\dot \Phi =V_{\mathrm{out}} +  V_{\mathrm{in}}, \hspace{0.3cm} I_{\rm wg} = \frac{\dot \Phi - 2 V_{\mathrm{in}}}{Z_0}. \label{eq:galvanic_coupling}
\end{equation}
Using this, we recast \cref{eq:CurrentNodeEquationDirectCoupling} to
\begin{equation}
    C \ddot \Phi + \frac{\Phi}{L} + \frac{  \dot \Phi - 2 V_{\rm in}}{Z_0} = 0. \label{eq:EOMDirectCoupling}
\end{equation}
We Fourier transform this equation and find that
\begin{equation}
    \tilde{\Phi} = \frac{2L}{\omega^2 L C Z_0 - i \omega L - Z_0} \tilde{V}_{\mathrm{in}},
\end{equation}
where $\tilde{\mathcal{O}}$ denotes the Fourier transform of $\mathcal{O}$ to the frequency domain. Noting that $\tilde{V}_{\mathrm{out}} = i \omega \tilde{\Phi} - \tilde{V}_{\mathrm{in}}$, we find
\begin{equation}
    \frac{\tilde{V}_{\mathrm{out}}}{\tilde{V}_{\mathrm{in}}} =  -1 + \frac{2 i \omega L }{\omega^2 L C Z_0 - i \omega L - Z_0},  
\end{equation}
which is exactly the same reflection coefficient $\Gamma$ in \cref{eq:ReflectionCoefficientGalvanicCoupling}. 

We note that the Fourier method only works for EOMs that are linear. For nonlinear EOMs, we substitute $V_\mathrm{in} = A \cos (\omega t)$ and solve numerically for $\Phi$. In a long time limit when $\Phi$ reaches its steady state, we compute $V_\mathrm{out} = \dot \Phi - V_{\mathrm{in}}$ and the reflection coefficient $\Gamma$.  

\subsubsection{Capacitive coupling}

We consider the circuit in \cref{fig:LCCoupledToWaveguide}b, where a parallel LC circuit is coupled capacitively to a waveguide with a coupling capacitance $C_C$. We compute the reflection coefficient in this circuit via the load impedance, which is given by
\begin{equation}
    Z_{\mathrm{L}} \equiv Z_{C_C} + Z_{LC} = \frac{1}{i \omega C_C} +   \frac{i \omega L}{1- \omega^2 L C}.
\end{equation}
By this, the reflection coefficient is
\begin{equation}
    \Gamma \equiv \frac{Z_{\mathrm{L}} - Z_0}{Z_{\mathrm{L}} + Z_0} = \frac{\frac{1}{i \omega C_C} + \frac{i \omega L}{1-\omega^2 L C} - Z_0 }{\frac{1}{i \omega C} + \frac{i \omega L}{1-\omega^2 L C} + Z_0}. \label{eq:ReflectionCoefficientCapacitiveCoupling}
\end{equation}

We derive the EOMs of the circuit in \cref{fig:LCCoupledToWaveguide}b. Kirchhoff's current law at its two nodes yields
\begin{eqnarray}
     I_C + I_L   + I_{C_C} &=& 0 , \\ \label{eq:CurrentNodeEquationCapacitiveCoupling1}
     I_{\rm wg} - I_{C_C} &=& 0, \label{eq:CurrentNodeEquationCapacitiveCoupling2}
\end{eqnarray}
where $I_C = C \ddot \Phi$ and $I_L = \Phi/L$. We also have
\begin{equation}
 I_{C_C} = C_C (\ddot \Phi - \dot V_{\rm wg}), \hspace{0.3cm} I_{\rm wg} = \frac{V_{\rm wg} - 2V_{\mathrm{in}}}{Z_0}, 
\end{equation}
by which we  recast eqs.\ \eqref{eq:CurrentNodeEquationCapacitiveCoupling1} and \eqref{eq:CurrentNodeEquationCapacitiveCoupling2} to
\begin{eqnarray}
     C \ddot \Phi + \frac{\Phi}{L} + C_C (\ddot \Phi - \dot V_{\rm wg}) &=& 0, \label{eq:EOMCapacitiveCoupling1} \\
    \frac{V_{\rm wg} - 2V_{\mathrm{in}}}{Z_0} - C_C (\ddot \Phi - \dot V_{\rm wg}) &=& 0. \label{eq:EOMCapacitiveCoupling2}
\end{eqnarray} 
We Fourier transform eqs.\ \eqref{eq:EOMCapacitiveCoupling1} and \eqref{eq:EOMCapacitiveCoupling2}  to solve for $\tilde{\Phi}$ and $\tilde{V}_{\rm wg}$ in terms of $\tilde{V}_{\mathrm{in}}$ and find that
\begin{equation}
    \tilde{V}_{\rm wg} =   \frac{2i - 2 i \omega^2 L (C + C_C)}{\omega^3 L C C_C Z_0 - \omega^2 L (C+C_C) - \omega C_C Z_0 + i}  \tilde{V}_{\mathrm{in}}.
\end{equation}
Noting that $\tilde{V}_{\mathrm{out}} =  \tilde{V}_{\rm wg} - \tilde{V}_{\mathrm{in}}$  we find
\begin{equation}
 \frac{\tilde{V}_{\mathrm{out}}}{\tilde{V}_{\mathrm{in}}} =  -1+ \frac{2i - 2 i \omega^2 L (C + C_C)}{\omega^3 L C C_C Z_0 - \omega^2 L (C+C_C) - \omega C_C Z_0 + i},
\end{equation}
which is exactly the same reflection coefficient $\Gamma$  in \cref{eq:ReflectionCoefficientCapacitiveCoupling}. 

In the limit $C_C \to \infty$, we recover galvanic coupling from capacitive coupling, where the reflection coefficient $\Gamma$ in \cref{eq:ReflectionCoefficientCapacitiveCoupling} is reduced exactly to the one in Eq.\ \eqref{eq:ReflectionCoefficientGalvanicCoupling}. In some electrical circuits
the limit \mbox{$C\to \infty$} or \mbox{$C\to 0$} must be taken with care due to potentially singular Lagrangians \cite{RymarzPRX23}. 

\subsection{SG equation without waveguide coupling}

\begin{figure}[t!]
    \centering
    \includegraphics[width = 0.45\textwidth]{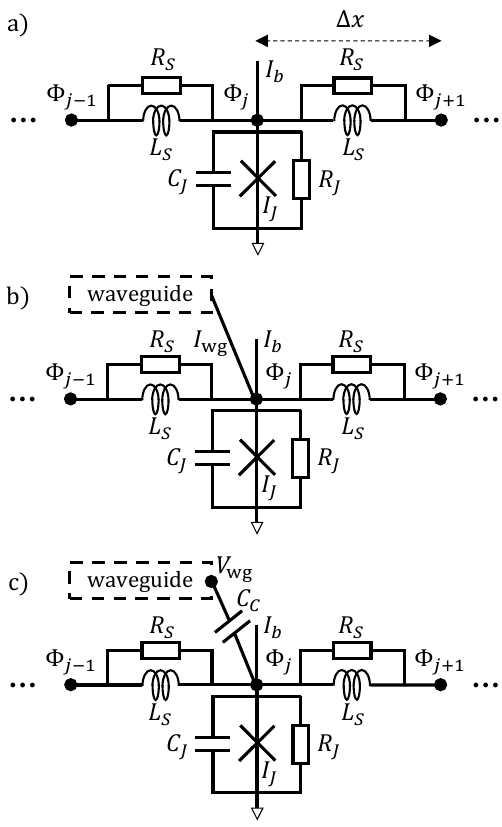}
    \caption{a) Discretisation of a long Josephson junction into discrete nodes equally separated by $\Delta x$. Each node is associated to a node flux  $ \Phi_j $, driven by a current bias $I_b$, and connected to the ground by a small Josephson junction in the RSCJ model \cite{TinkhamBook2004} with a critical current $I_J$, a parasitic capacitance $C_J$, and a shunted resistance $R_J$. Subsequent nodes are connected to each other by a surface inductance $L_S$ and a surface resistance $R_S$. b) and c) Same as panel a) but node $j$ is galvanically or capacitively coupled to a waveguide, respectively. The lumped-element model of a waveguide is shown in \cref{fig:LCCoupledToWaveguide}.  }
    \label{fig:DiscreteLongJunction}
\end{figure}

Here we derive the SG equation of a LJJ that includes external current bias and dissipation \cite{fedorov2013fluxon,MeckbachPhdThesis2013}. To this end, we discretise the LJJ into discrete unit cells of length $\Delta x$, each of which is associated to a node flux, as shown in \cref{fig:DiscreteLongJunction}a. Each node is connected to the ground by a small Josephson junction in the RSCJ model \cite{TinkhamBook2004} with a Josephson critical current $I_J$, a parasitic capacitance $C_J$, and a shunted resistance $R_J$. Two consecutive nodes are coupled to each other by a surface inductance $L_S$ and a surface resistance $R_S$.

We apply Kirchhoff's current law to node $j$ with its node flux $\Phi_j$ and obtain
\begin{eqnarray}
   && C_J \ddot \Phi_j + I_J \sin \left( \frac{2\pi}{\Phi_0} \Phi_j  \right) - \frac{\Phi_{j-1}-2 \Phi_j + \Phi_{j+1} }{L_S} \nonumber \\
   && + \frac{\dot \Phi_j}{R_J} - \frac{\dot \Phi_{j-1}-2\dot \Phi_j + \dot \Phi_{j+1}}{R_S} = I_b. \label{eq:SGequation1}
\end{eqnarray}
We define per-unit-length quantities: $C_J = c_J \Delta x$, $I_J = i_J \Delta x$, $L_S=l_S \Delta x$, $R_S = r_S \Delta x$, and $I_b = i_b \Delta x$ with $c_J$, $i_J$, $l_S$, $r_S$, and $i_b$ the junction capacitance, the critical junction current density, the surface inductance, the surface resistance, and the bias current per unit length, respectively. We substitute these back to Eq.\ \eqref{eq:SGequation1}, divide both of its sides by $\Delta x$, and convert the flux to the phase $\Phi_j = (\Phi_0/2\pi) \phi_j$ to obtain
\begin{eqnarray}
   &&  \frac{\Phi_0 c_J}{2\pi} \ddot \phi_j + i_J \sin (\phi_j) - \frac{\Phi_0}{2\pi l_S} \frac{\phi_{j-1}-2\phi_j + \phi_{j+1}}{(\Delta x)^2} \nonumber \\
   && + \frac{\Phi_0}{2\pi \tilde R_J } \dot \phi_j - \frac{\Phi_0}{2\pi r_S} \frac{\dot \phi_{j-1} - 2\dot \phi_j + \dot \phi_{j+1}}{(\Delta x)^2} = i_b, \label{eq:SGequation2}
\end{eqnarray}
where $\tilde R_J = R_J \Delta x$ is the sheet resistance (over a unit length). 
Dividing both sides of Eq.\ \eqref{eq:SGequation2} by $i_J$,  normalise time $t\equiv \textsf{t} \omega_p^{-1}$ with  $\omega_p = \sqrt{2\pi i_J/(\Phi_0 c_J)}$ the plasma frequency and position $x \equiv \textsf{x} \lambda_J$ with $\lambda_J = \sqrt{\Phi_0/(2\pi l_S i_J)}$ the Josephson penetration depth, and taking the limit $\Delta \textsf{x} \to 0$, we achieve the final form of the SG equation
\begin{equation}
    \phi_{\textsf{tt}} + \sin(\phi) - \phi_{\textsf{xx}} + \textsf{g} \phi_{\textsf{t}} - \textsf{p} \phi_{\textsf{xxt}} = \textsf{i}_b, \label{eq:SG_after_variable_normalisation}
\end{equation}
where $\textsf{i}_b = i_b/i_J$,  $\textsf{g} = \sqrt{\Phi_0/2\pi c_J \tilde R_J^2 i_J}$, and $\textsf{p} = \sqrt{2\pi l_s^2 i_J/\Phi_0 r_S^2 c_J} $.

\subsection{SG equation with galvanic waveguide coupling}

We assume coupling node $j$ galvanically to a waveguide associated with a current $I_{\mathrm{wg}}$, as shown in \cref{fig:DiscreteLongJunction}b. 
The equation of motion for the flux $\Phi_j$ at node $j$ is the same as Eq.\ \eqref{eq:SGequation1} except that the RHS includes also $I_{\mathrm{wg}}$
\begin{eqnarray}
   && C_J \ddot \Phi_j + I_J \sin \left( \frac{2\pi}{\Phi_0} \Phi_j  \right) - \frac{\Phi_{j-1}-2 \Phi_j + \Phi_{j+1} }{L_S} \nonumber \\
   && + \frac{\dot \Phi_j}{R_J} - \frac{\dot \Phi_{j-1}-2\dot \Phi_j + \dot \Phi_{j+1}}{R_S} = I_b - I_{\mathrm{wg}}, \label{eq:SGequationDirectCoupling1}
\end{eqnarray}
where (see Eq. \eqref{eq:galvanic_coupling})
\begin{eqnarray}
    I_{\mathrm{wg}} &=& \frac{\dot \Phi_j - 2 V_{\mathrm{in},j}}{Z_0}. \\
    \dot \Phi_j &=& V_{\mathrm{in},j} + V_{\mathrm{out},j}. \label{eq:InputOutputNodej}
\end{eqnarray}
Converting relevant quantities to their per-unit-length counterparts and the flux $\Phi_j$ to the phase $\phi_j$ and dividing all terms by $\Delta x$ as done in Eq.\ \eqref{eq:SGequation2}, we rewrite Eq.\ \eqref{eq:SGequationDirectCoupling1} to
\begin{eqnarray}
   &&  \frac{\Phi_0 c_J}{2\pi} \ddot \phi_j + i_J \sin (\phi_j) - \frac{\Phi_0}{2\pi l_S} \frac{\phi_{j-1}-2\phi_j + \phi_{j+1}}{(\Delta x)^2} \nonumber \\
   && + \frac{\Phi_0}{2\pi \tilde R_J } \dot \phi_j - \frac{\Phi_0}{2\pi r_S} \frac{\dot \phi_{j-1} - 2\dot \phi_j + \dot \phi_{j+1}}{(\Delta x)^2} \nonumber \\
   &=& i_b + \frac{\Phi_0}{2\pi \Delta x} \frac{2 \omega_p \textsf{V}_{\mathrm{in},j} - \dot \phi_j}{Z_0} , \label{eq:SGequationDirectCoupling2}
\end{eqnarray}
where $\textsf{V}_{\mathrm{in},j} 
 = V_{\mathrm{in},j}/ ( \Phi_0 \omega_p/2\pi) $ is  the dimensionless input voltage. 
We normalise $x$ to $\lambda_J$ and $t$ to $\omega_p^{-1}$, divide all terms by $i_J$, use the identity  $\Phi_0 \omega_p/(2\pi \lambda_J i_J) = \sqrt{l_S/c_J} \equiv Z_{\mathrm{LJJ}}$ with $Z_{\rm LJJ}$ the characteristic LJJ impedance, and take the continuum limit $\Delta \textsf{x} \to 0$ to 
recast eqs.\ \eqref{eq:SGequationDirectCoupling2} and \eqref{eq:InputOutputNodej} to
\begin{eqnarray}
&& \phi_{\textsf{tt}} + \sin(\phi) - \phi_{\textsf{xx}} + \textsf{g} \phi_{\textsf{t}} - \textsf{p} \phi_{\textsf{xxt}} \nonumber \\
&=& \textsf{i}_b + \textsf{z} \delta (\textsf{x} - \textsf{x}_j)  (2 \textsf{V}_{\mathrm{in},j}-\phi_{\textsf{t}}), \label{eq:SGequationDirectCoupling3}
\end{eqnarray}
and 
\begin{equation}
 \phi_{\textsf t} (\textsf{x}_j) = \textsf{V}_{\mathrm{in},j} + \textsf{V}_{\mathrm{out},j},
\end{equation}
where $\textsf{z}=Z_{\rm LJJ}/Z_0$. For $n_{\rm wg}$ galvanic waveguide couplings, we replace the last term in the RHS of Eq.\ \eqref{eq:SGequationDirectCoupling3} by $\sum_j^{n_{\rm wg}} \textsf{z} \delta (\textsf{x} - \textsf{x}_j)  (2 \textsf{V}_{\mathrm{in},j}-\phi_{\textsf{t}})$.
This explains the coupling form $\mathcal C_{\rm x} [\phi]$ and the input-output relation for galvanic waveguide coupling  we use in the main text.

The impedance ratio $\textsf{z}$ in Eq.\ \eqref{eq:SGequationDirectCoupling3} represents the coupling strength in scattering calculations. Therefore, it accounts for the effect of impedance mismatch between the LJJ and waveguide ports, i.e., large mismatch results in a small circulation bandwidth.

\subsection{SG equation with capacitive waveguide coupling}
We assume coupling node $j$ capacitively to a waveguide, as shown in \cref{fig:DiscreteLongJunction}c.  
The EOMs for the phase $\phi_j$ and the normalised waveguide voltage $\textsf{V}_{\mathrm{wg},j} = V_{\mathrm{wg},j}/ ( \Phi_0 \omega_p/2\pi) $ (which are similar to Eqs.\ \eqref{eq:EOMCapacitiveCoupling1} and \eqref{eq:EOMCapacitiveCoupling2}) include  
\begin{eqnarray}
  &&  \frac{\Phi_0 c_J}{2\pi} \ddot \phi_j + i_J \sin (\phi_j) - \frac{\Phi_0}{2\pi l_S} \frac{\phi_{j-1}-2\phi_j + \phi_{j+1}}{(\Delta x)^2} \nonumber \\
  && + \frac{\Phi_0}{2\pi \tilde R_J } \dot \phi_j - \frac{\Phi_0}{2\pi r_S} \frac{\dot \phi_{j-1} - 2\dot \phi_j + \dot \phi_{j+1}}{(\Delta x)^2} \nonumber \\
  &=& i_b + \frac{\Phi_0 C_C} {2\pi \Delta x} ( \omega_p \dot {\textsf{V}}_{\mathrm{wg},j} - \ddot \phi_j). \label{eq:CapacitiveCouplingNodej1}
\end{eqnarray}
and 
\begin{equation}
\frac{\Phi_0 \omega_p}{2\pi Z_0} (\textsf{V}_{\mathrm{wg},j} - 2 \textsf{V}_{\mathrm{in},j}) + \frac{\Phi_0 C_C} {2\pi } ( \omega_p \dot {\textsf{V}}_{\mathrm{wg},j} - \ddot \phi_j)  = 0, \label{eq:CapacitiveCouplingNodej2}
\end{equation}
We normalise $x$ to $\lambda_J$ and $t$ to $\omega_p^{-1}$, divide all terms by $i_J$, and take the continuum limit $\Delta \textsf{x} \to 0$. This recasts Eqs. \eqref{eq:CapacitiveCouplingNodej1} and \eqref{eq:CapacitiveCouplingNodej2} to
\begin{eqnarray}
&& \phi_{\textsf{tt}} + \sin(\phi) - \phi_{\textsf{xx}} + \textsf{g} \phi_{\textsf{t}} - \textsf{p} \phi_{\textsf{xxt}} \nonumber \\
&=& \textsf{i}_b + \delta (\textsf{x} - \textsf{x}_j) \textsf{g}_C ( \partial_{\textsf t} \textsf{V}_{j}-\partial_{\textsf{tt}} \phi), \label{eq:CapacitiveCouplingNodej3}
\end{eqnarray}
and 
\begin{equation}
\textsf{p}_C (\textsf V_j - 2 \textsf V_{\mathrm{in},j}) + (\partial_{\textsf t} \textsf V_j - \partial_{\textsf{tt}} \phi_j) = 0, \label{eq:CapacitiveCouplingNodeVj3}
\end{equation}
where $\textsf{g}_{C} = Z_{\rm LJJ}/Z_{C}$ and $\textsf p_C = {Z_C}/{Z_0}$ with $Z_{C} = 1/(\omega_p C_C)$. The input-output relation in this case is
\begin{equation}
\textsf{V}_{\mathrm{wg},j} = \textsf{V}_{\mathrm{in},j} + \textsf{V}_{\mathrm{out}, j}. 
\end{equation}
For $n_{\rm wg}$ capacitive waveguide couplings, we replace the last term in the RHS of Eq.\ \eqref{eq:CapacitiveCouplingNodej3} by $\sum_j^{n_{\rm wg}} \delta (\textsf{x} - \textsf{x}_j) \textsf{g}_C ( \partial_{\textsf t} \textsf{V}_{j}-\partial_{\textsf{tt}} \phi)$ and there are $n_{\rm wg}$ equations in the form of \cref{eq:CapacitiveCouplingNodeVj3}. Also, it is straightforward to verify that in the limit $C_C \to \infty$ the EOMs and the input-output relation derived above recover those of galvanic waveguide coupling. Below \cref{eq:InputOutputRelation} in the main text, we have described how to compute the $S$-matrix elements $S_{jk}$ from numerically solving the SG equation. 

The EOMs above can be converted to Lagrangians, which form the basis for circuit quantisation to consider quantum effects. Moving fluxons - the source of nonreciprocity in our system - might be analysed via the recently developed quantisation procedures for nonreciprocal circuits \cite{rymarz2018quantum,ParraRodriguezPRB19,EgusquizaPRB22,RymarzPRX21,
ParraRodriguez2022canonical,rymarz2023nonreciprocal}. 

\section{Additional numerical simulations}

\subsection{Scattering performance versus the fluxon number}

\begin{figure}[h]
    \centering
    \includegraphics{Optimal_Smatrix_versus_fluxon_number_26Sep25.pdf}
    \caption{Optimal $S$-matrix elements $S_{j1}$ ($j=1,2,3$) as functions of the fluxon number $n$. Simulation parameters are the same as those used in the main text.}
    \label{fig:Smatrix_vs_fluxon_number}
\end{figure}

We numerically examine the (optimal) scattering performance of our proposed circulator with respect to different fluxon numbers $n=4-15$, as shown in \cref{fig:Smatrix_vs_fluxon_number}. We observe that $S_{j1}$ ($j=1,2,3$) essentially remain unchanged for $n=7-15$, so in theory there would be no significant difference in circulation performance when operating our device at these different fluxon numbers.  We also find that the scattering performance starts diminishing for larger fluxon numbers $n\geq19$. In this case, circulation conditions require large bias currents and high fluxon velocities that lead to  relativistic effects.

In practice, there are several considerations to take into account when choosing the fluxon number $n$. First, for certain device parameters the fluxon number $n$ is chosen such that the operating frequency belongs to the target frequency range, such as $6-8$ GHz in superconducting qubits.  Second, the fluxon number $n$ should also be optimised for the fluxon-insertion approach used in experiments. For example, if using DC current injection \cite{Ustinov02}, small fluxon numbers are preferred to avoid excessively large injection current.

\subsection{Galvanic coupling versus capacitive coupling}

\begin{figure}[t]
    \centering
    \includegraphics{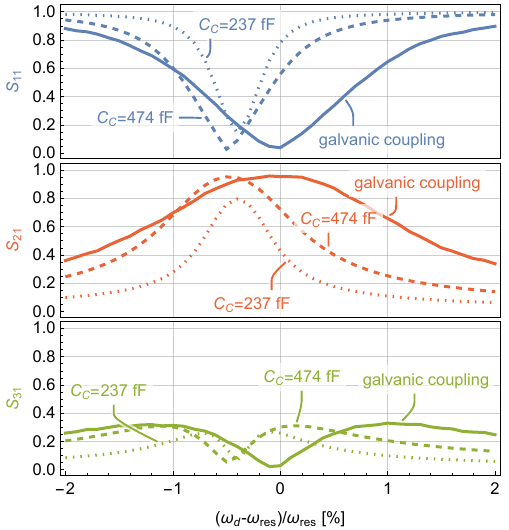}
    \caption{$S$-matrix elements $S_{j1}$ ($j=1,2,3$) as functions of the relative drive frequency detuning from the resonant value, $(\omega_d - \omega_{\rm res})/\omega_{\rm res}$, for galvanic coupling (solid lines) and for capacitive coupling with $C_C = 474\, \rm fF$ (dashed lines) and $C_C = 237\, \rm fF$ (dotted lines). Simulation parameters are the same as those used in the main text.}
    \label{fig:Galvanic_versus_capacitive}
\end{figure}

One can use both galvanic and capacitive waveguide couplings to couple our LJJ circulator to the external waveguide ports. As seen in eqs.\ \eqref{eq:CapacitiveCouplingNodej3} and \eqref{eq:CapacitiveCouplingNodeVj3}, capacitive waveguide coupling is characterised by the coupling capacitance $C_C$ and its impedance $Z_C=1/\omega_p C_C$. In \cref{fig:Galvanic_versus_capacitive}, we analyse the scattering spectrum versus the relative drive detuning to the LJJ resonant frequency, $(\omega_d - \omega_{\rm res})/\omega_{\rm res}$, for galvanic coupling (solid lines) and for capacitive coupling with $C_C = 474\, \rm fF$ (dashed lines) and $C_C = 237 \, \rm fF$ (dotted lines). These values of $C_C$ correspond to $Z_C = \{10 \Omega, 20 \Omega\}$, using the same simulation parameters as in the main text.  

We observe in \cref{fig:Galvanic_versus_capacitive} that the scattering spectrum for galvanic coupling is essentially symmetric around the zero drive detuning point, while those for capacitive coupling $C_C = \{ 474\, \mathrm{fF}, 237\, \mathrm{fF} \}$ are shifted towards negative detuning points. The best $S_{j1}$ for galvanic coupling and for capacitive coupling with $C_C =  474\, \mathrm{fF} $ are relatively the same, close the ideal circulation with $\{S_{11},S_{21}, S_{31} \} \sim \{0,1,0\}$. For $C_C =  237\, \mathrm{fF}$ it shows a decline in circulation performance with $\{S_{11},S_{21}, S_{31} \} \sim \{0.2,0.8,0.1\}$. 
More importantly, when changing from galvanic coupling to capacitive coupling $C_C = \{ 474\, \mathrm{fF}, 237\, \mathrm{fF} \}$ the circulation bandwidth is significantly reduced. Increasing $C_C$ to a few pF, we find that capacitive coupling yields a scattering performance comparable to that of galvanic coupling in terms of both $S$-matrix elements and bandwidth. 
 
We also note that the scattering spectra in \cref{fig:Galvanic_versus_capacitive} are computed after optimising the current bias $\textsf{i}_b$ for galvanic coupling and for capacitive coupling with $C_C = 474\, \mathrm{fF}, $ and $C_C = 237\, \mathrm{fF} $. The corresponding optimal values are \mbox{$3\times 10^{-4}, 2.25\times 10^{-6}$}, and $  2.5\times 10^{-6}$. This shows another key difference in the operation of the LJJ circulator with galvanic versus capacitive waveguide coupling.

Viola \textit{et al} \cite{PhysRevX.4.021019} in analysing Hall-effect-based nonreciprocal devices has pointed out that galvanic coupling could introduce unwanted Ohmic heating at the contact points, which can be avoided using reactive coupling such as capacitive coupling. In a future work, we will take into account the effect of this Ohmic dissipation in our galvanically coupled circulator device.

\end{document}